\documentclass[runningheads]{llncs}
\usepackage[T1]{fontenc}
\usepackage{xcolor, xspace}
\usepackage{listings}
\usepackage{color,soul}
\usepackage{mathtools}

\usepackage{amsthm}

\usepackage{float}
\usepackage{subfig}
\usepackage{graphicx}
\usepackage{calc}

\newcommand{\citet}[1]{\cite{#1}}
\newcommand{\citep}[1]{\cite{#1}}

\newfloat{lstfloat}{htbp}{lop}
\floatname{lstfloat}{Listing}

\definecolor{dkgreen}{rgb}{0,0.6,0}
\definecolor{gray}{rgb}{0.5,0.5,0.5}
\definecolor{mauve}{rgb}{0.58,0,0.82}

\definecolor{dkgreen}{rgb}{0,0.6,0}
\definecolor{ltblue}{rgb}{0,0.4,0.4}
\definecolor{dkviolet}{rgb}{0.3,0,0.5}
\definecolor{dkblue}{rgb}{0.2,0.2,0.6}
\definecolor{dkred}{rgb}{0.5,0.0,0.13}
\lstdefinelanguage{Coq}{ 
    mathescape=true,
    texcl=false, 
    morekeywords=[1]{Section, Module, End, Require, Import, Export,
        Variable, Variables, Parameter, Parameters, Axiom, Hypothesis,
        Hypotheses, Notation, Local, Tactic, Reserved, Scope, Open, Close,
        Bind, Delimit, Definition, Let, Ltac, Fixpoint, CoFixpoint, Add,
        Morphism, Relation, Implicit, Arguments, Unset, Contextual,
        Strict, Prenex, Implicits, Inductive, CoInductive, Record,
        Structure, Canonical, Coercion, Context, Class, Global, Instance,
        Program, Infix, Theorem, Lemma, Corollary, Proposition, Fact,
        Remark, Example, Proof, Goal, Save, Qed, Defined, Hint, Resolve,
        Rewrite, View, Search, Show, Print, Printing, All, Eval, Check,
        Projections, inside, outside, Def},
    morekeywords=[2]{forall, exists, exists2, fun, fix, cofix, struct,
        match, with, end, as, in, return, let, if, is, then, else, for, of,
        nosimpl, when},
    morekeywords=[3]{Type, Prop, Set, true, false, option},
    morekeywords=[4]{pose, set, move, case, elim, apply, clear, hnf,
        intro, intros, generalize, rename, pattern, after, destruct,
        induction, using, refine, inversion, injection, rewrite, congr,
        unlock, compute, ring, field, fourier, replace, fold, unfold,
        change, cutrewrite, simpl, have, suff, wlog, suffices, without,
        loss, nat_norm, assert, cut, trivial, revert, bool_congr, nat_congr,
        symmetry, transitivity, auto, split, left, right, autorewrite},
    morekeywords=[5]{by, done, exact, reflexivity, tauto, romega, omega,
        assumption, solve, contradiction, discriminate},
    morekeywords=[6]{do, last, first, try, idtac, repeat},
    morecomment=[s]{(*}{*)},
    showstringspaces=false,
    tabsize=3,
    extendedchars=false,
    sensitive=true,
    breaklines=false,
    basicstyle=\small,
    captionpos=b,
    columns=[l]flexible,
    identifierstyle={\ttfamily\color{black}},
    keywordstyle=[1]{\ttfamily\color{dkviolet}},
    keywordstyle=[2]{\ttfamily\color{dkgreen}},
    keywordstyle=[3]{\ttfamily\color{ltblue}},
    keywordstyle=[4]{\ttfamily\color{dkblue}},
    keywordstyle=[5]{\ttfamily\color{dkred}},
    stringstyle=\ttfamily,
    commentstyle={\ttfamily\color{dkgreen}},
    literate=
    {\\forall}{{\color{dkgreen}{$\forall\;$}}}1
    {\\exists}{{$\exists\;$}}1
    {<-}{{$\leftarrow\;$}}1
    {=>}{{$\Rightarrow\;$}}1
    {==}{{\code{==}\;}}1
    {==>}{{\code{==>}\;}}1
    {->}{{$\rightarrow\;$}}1
    {<->}{{$\leftrightarrow\;$}}1
    {<==}{{$\leq\;$}}1
    {\#}{{$^\star$}}1 
    {\\o}{{$\circ\;$}}1 
    {\@}{{$\cdot$}}1 
    {\/\\}{{$\wedge\;$}}1
    {\\\/}{{$\vee\;$}}1
    {++}{{\code{++}}}1
    {~}{{$\sim$}}1
    {\@\@}{{$@$}}1
    {\\mapsto}{{$\mapsto\;$}}1
    {\\hline}{{\rule{\linewidth}{0.5pt}}}1
}[keywords,comments,strings]

\newlength\myheight
\newlength\mydepth
\settototalheight\myheight{Xygp}
\newcommand*\inlinegraphics[1]{%
\raisebox{-0.15\baselineskip}{%
\includegraphics[
height=0.8\baselineskip,
width=0.8\baselineskip,
keepaspectratio,
]{#1}%
}%
}

\newcommand{\coqverified}[0]{\inlinegraphics{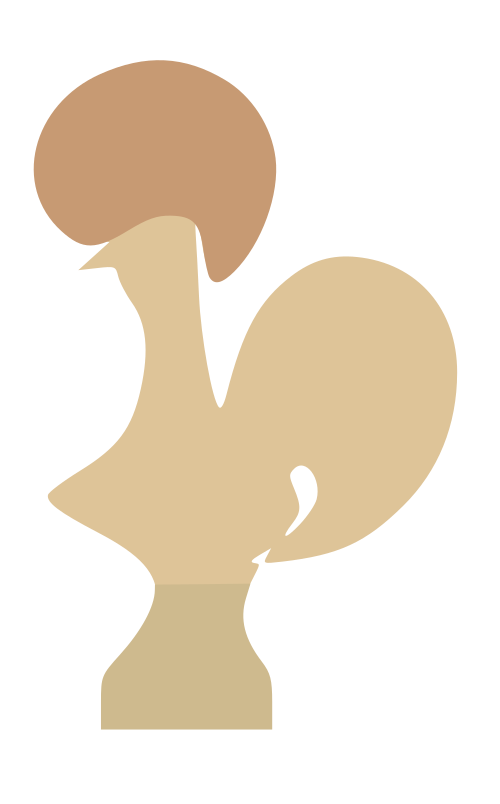}}

\newtheoremstyle{ndefinition} % name
    {\topsep}                    % Space above
    {.8em}                    % Space below
    {}                   % Body font
    {}                           % Indent amount
    {\bf \scshape}                   % Theorem head font
    {}              % Punctuation after theorem head
    {.7em}                       % Space after theorem head
    {}  % Theorem head spec (can be left empty, meaning ‘normal’)
\theoremstyle{ndefinition} 
\newtheorem{defn}{Definition}

\newtheoremstyle{cdefinition} % name
    {\topsep}                    % Space above
    {.8em}                    % Space below
    {}                   % Body font
    {}                           % Indent amount
    {\bf \scshape}                   % Theorem head font
    { \coqverified}              % Punctuation after theorem head
    {\baselineskip}                       % Space after theorem head
    {}  % Theorem head spec (can be left empty, meaning ‘normal’)

\theoremstyle{cdefinition} 
\newtheorem{cthm}{Theorem}

\newtheorem{cexample}{Example}

\lstdefinelanguage{MSWasm}{ 
    mathescape=true,
    texcl=false, 
    classoffset=1,
    morekeywords={i32,f32,i64,f64, handle},
    keywordstyle=\color{ltblue},
    classoffset=2,
    morekeywords={add,const, if,  sub, TO_eqz, testop, eqz,
    module, func, param, result, global, 
             global.get, mut, global.set, export, import, memory, data, 
             local.get, local.set, local.tee, elem, table, call,call_indirect, type,
             segalloc, segstore, segload},
    keywordstyle=\color{blue},
    morecomment=[s]{;;}{;;},
    alsoletter={.},
    showstringspaces=false,
    tabsize=3,
    extendedchars=false,
    sensitive=true,
    breaklines=false,
    basicstyle={\small\ttfamily},
    captionpos=b,
    columns=[l]flexible,
    identifierstyle={\ttfamily\color{black}},
    keywordstyle=\color{blue},
    commentstyle=\color{dkgreen},
    stringstyle=\ttfamily,
    commentstyle={\ttfamily\color{dkgreen}},
}[keywords,comments,strings]
\lstdefinelanguage{MSWasmNoMath}{ 
    language=C++,
    aboveskip=3mm,
    belowskip=3mm,
    showstringspaces=false,
    columns=flexible,
    basicstyle={\small\ttfamily},
    numbers=none,
    alsoletter={.$ $},
    numberstyle=\tiny\color{gray},
    classoffset=1,
    morekeywords={i32,f32,i64,f64, handle},
    keywordstyle=\color{ltblue},
    classoffset=2,
    morekeywords={add,const, if,  sub, TO_eqz, testop, eqz,
    module, func, param, result, global, 
             global.get, mut, global.set, export, import, memory, data, 
             local.get, local.set, local.tee, elem, table, call,call_indirect, type,
             segalloc, segstore, segload},
    keywordstyle=\color{blue},
    commentstyle=\color{dkgreen},
    stringstyle=\color{mauve},
    breaklines=true,
    breakatwhitespace=true,
    tabsize=3,
      numbers=left
}[keywords,comments,strings]

\lstdefinelanguage{CheckedC}{ frame=tb,
    language=C++,
    aboveskip=3mm,
    belowskip=3mm,
    showstringspaces=false,
    columns=flexible,
    basicstyle={\small\ttfamily},
    numbers=none,
    alsoletter={.},
    numberstyle=\tiny\color{gray},
    keywordstyle=\color{blue},
    classoffset=1,
    morekeywords={restrict, u32, uint8_t, size_t, ptr, array_ptr, nt_array_ptr, Handle, itype_for_any, itype},
    keywordstyle=\color{blue},
    classoffset=2,
    morekeywords={ count, byte_count},
    keywordstyle=\color{ltblue},
    commentstyle=\color{dkgreen},
    stringstyle=\color{mauve},
    breaklines=true,
    breakatwhitespace=true,
    tabsize=3,
      numbers=left
}[keywords,comments,strings]

\usepackage{authcomments}
\newcommenter{EDJ}{0.9, 0.8,1.0}
\newcommenter{john}{1.0, 0, 0}

\usepackage{hyperref}

\usepackage{cleveref}

\newcommand{\ctomswasm}{C2M\xspace}
\newcommand{\mswasmtocc}{M2C\xspace}
\newcommand{\checkedc}{Checked C\xspace}
\newcommand{\mswasm}{MSWASM\xspace}
\newcommand{\wasm}{WASM\xspace}

\newcommand{\aegis}{Aegis\xspace}

\begin{document}
\title{AEGIS: Towards Formalized and Practical Memory-Safe Execution of C programs via MSWASM}
\author{%
Shahram Esmaeilsabzali\inst{1} \and 
Arayi Khalatyan\inst{2}\thanks{This author was a co-op at the Waterloo Research Centre at the time this research was carried out.} \and
Zhijun Mo\inst{1} \and 
Sruthi Venkatanarayanan\inst{1} \and 
Shengjie Xu\inst{1}
}%
\institute{
Huawei Canada, Waterloo Research Centre\\
\email{shahram.esmaeilsabzali@huawei.com}\\
\email{zhijun.mo@huawei.com}\\
\email{sruthi.venkatanarayanan@huawei.com}\\
\email{shengjie.xu@huawei.com}
\and
University of Waterloo\\
\email{akhalaty@uwaterloo.ca}}

\maketitle              % typeset the header of the contribution
\begin{abstract}
Programs written in unsafe languages such as C are prone to memory safety errors, which can lead to program compromises and serious real-world security consequences. Recently, Memory-Safe WebAssembly~(MSWASM) is introduced as a general-purpose intermediate bytecode with built-in memory safety semantics. Programs written in C can be compiled into MSWASM to get complete memory safety protection. 

In this paper, we present our extensions on MSWASM, which improve its semantics and practicality. First, we formalize MSWASM semantics in Coq/Iris, extending it with inter-module interaction, showing that MSWASM provides fine-grained isolation guarantees analogous to WASM's coarse-grained isolation via linear memory. Second, we present \aegis, a system to adopt the memory safety of MSWASM for C programs in an interoperable way. \aegis pipeline generates \checkedc source code from MSWASM modules to enforce spatial memory safety. \checkedc is a recent binary-compatible extension of C which can provide guaranteed spatial safety. Our design allows \aegis to protect C programs that depend on legacy C libraries with no extra dependency and with low overhead.
\aegis pipeline incurs 67\% runtime overhead and near-zero memory overhead on PolyBenchC programs compared to native.
\end{abstract}

\section{Introduction}

Memory safety, which ensures that a pointer is valid and its dereference accesses a valid memory location, is of paramount importance in systems security. There is a huge amount of C code in modern software infrastructures. However, C semantics does not protect these codes from memory safety violations like buffer overflow or use-after-free. Therefore, a single out-of-bound pointer dereference can be, and has been, the root cause of critical security vulnerabilities. For example, the Heartbleed vulnerability~\cite{heartbleed} in 2014 would allow a remote attacker to dump server process memory stealthily, which can leak private keys and user data. There is a plethora of approaches to this problem~\cite{6547101}, including compiler instrumentation~\cite{sanitizers,mpxexplained,softbound}, fuzzing~\cite{fuzzing_comprehensive,fuzzing_survey_roadmap}, or rewriting C code into a memory safe languages, such as Rust~\cite{c2rust_aliasing,c2rust_ownership}. However, there is no silver bullet to the problem yet because there is a trade-off between the performance, practicality, and security guarantees of different approaches.

Recently, Memory-Safe WebAssembly(\mswasm) has introduced a new direction in memory safety enforcement for C programs~\cite{michael2023mswasm}. \mswasm is based on \wasm, an intermediate, portable bytecode format that C/C++ programs can be compiled into. \wasm programs can run efficiently in secure sandboxes (e.g., browsers). However, \wasm does not prevent memory safety violations within the program~\cite{michael2023mswasm}. By extending raw pointers into \mswasm \emph{handles}, which are unforgeable, integrity-protected fat pointers, \mswasm could provide complete memory safety for C. In addition, as an intermediate bytecode format, \mswasm allows the execution engine or source-to-source translator that implements it to decide the exact data layout of handles and the security policy to enforce for the same bytecode. This feature offers users the freedom to choose appropriate security-performance tradeoffs depending on a deployment scenario.

While the original \mswasm paper established the foundation for memory safety enforcement~\cite{michael2023mswasm}, there are still missing pieces for practical adoption. First, there is no mechanical formalization of the semantics of \mswasm  and mechanized proofs of its memory safety properties, both of which are necessary for ensuring practical safety and security. The memory safety guarantees of \mswasm, especially \textit{spatial safety}, which ensures lack of buffer overflow and underflow, has been proven informally, in the context of single-module semantics only. In particular, it does not deal with consequences of importing and exporting segments of memory via handles belonging to different \mswasm modules. As a result, it is not clear how to formulate the isolation guarantees for real-world \mswasm modules with external dependencies. Second, the \mswasm prototype cannot generate code that depends on legacy C libraries, because compiling a C program into \mswasm requires \emph{all} dependent code to be compiled with the same toolchain; it does not allow for partial compilation of a C library or file. This significantly limits the usefulness of \mswasm for existing systems. 

In this paper, we present \aegis, which extends \mswasm both formally and practically in the following two ways:

\paragraph{Formalism.}
We present a full formalization of \mswasm semantics in Coq/Iris~\cite{jung2016higher}, and prove \emph{local handle isolation} for \mswasm, a fine-grained notion of isolation comparable to the coarse-grained \wasm isolation~\cite{rao2023iris}, proven recently. We extend the recent \wasm semantic formalization and theorems~\cite{rao2023iris,watt2021two} to include all features of \mswasm, specifically, the inter-module interaction. We also provide a proof of our compilation from \mswasm to \checkedc, thereby guaranteeing the spatial safety of our generated
 \checkedc code.

\paragraph{\aegis Pipeline.}
We present our toolchain for translating an \mswasm module to efficient, spatially-safe \checkedc code. \checkedc~\cite{ruef2017checked} is a binary-compatible extension of C that introduces a minimal type system to annotate pointers with bound information (referred to as \emph{checked} pointers). The \checkedc compiler either reasons that a checked pointer dereference is safe or adds dynamic bound checking, which can potentially be removed by compiler optimizations. The compilation pipeline involves (1) compiling the original C programs into \mswasm, (2) running a ``\mswasm to \checkedc'' (\mswasmtocc) code generator to emit \checkedc sources, and (3) compiling and linking the generated \checkedc sources with legacy C code to produce the executables. Incorporating \mswasm as the intermediate bytecode format simplifies formalization of spatial memory safety over the entire compilation process, especially when the final executable contains code from multiple parties. The \mswasmtocc code generator performs \textit{type-preserving} compilation~\cite{morrisett1999system}, where well-typed \mswasm code is translated into well-typed \checkedc code. To bridge the difference in pointer representation between \mswasm-protected code and legacy C code when they call functions in each other, the code generator emits \emph{wrapper functions} that convert arguments and return values for function calls. In addition, we implemented several features missing from \mswasm prototype in \aegis, including support for global variables and \emph{slice} instructions for spatial safety on stack. Lastly, we prove our translation algorithm produces spatially-safe \checkedc code.

We have benchmarked \aegis pipeline against PolyBenchC and compared the overhead against Rust backend of \mswasm. Our runtime overhead against native C code is 67\% on average, with memory overhead being negligible. The Rust implementation of \mswasm has runtime overhead of 96\% with a massive memory usage overhead compared to \aegis.

Our contribution in this paper is twofold:
\begin{itemize}
\item \textbf{Semantic formalization of \mswasm in Coq/Iris:} We mechanize \mswasm semantics in Coq/Iris. We state a theorem that characterizes the isolation guarantees and pointer integrity of \mswasm. %Our theorem establishes the fine-grained isolation guarantees of \mswasm based on handles and segment memory, analogous to coarse-grained linear memory isolation of \wasm. 
We also prove our translation from \mswasm to \checkedc produces spatially-safe code.

\item \textbf{Spatial memory safety for C programs with \mswasm:}  We present a toolchain to protect C programs via \mswasm and \checkedc.%It translates MSWASM modules compiled from C programs to \checkedc source automatically for spatial safety enforcement.

We make our tools, source code, and formalizations all available.\footnote{https://doi.org/10.5281/zenodo.14289850}

\end{itemize}    

The remainder of the paper is organized as follows.
Section~\ref{sec:bg} provides background on \mswasm and Checked C.
Section~\ref{sec:formal} presents our formalization of MSWASM. 
Section~\ref{sec:impl} discusses the \aegis pipeline, including the overall design and the implementation based on the LLVM and WABT fork for \mswasm.
Section~\ref{sec:eval} contains the evaluation for our implementation.
Section~\ref{sec:related} discusses related work in memory safety and relevant formal proofs.
Finally, Section~\ref{sec:conclusion} concludes the paper.

\section{Background}\label{sec:bg}

\subsection{WASM and MSWASM}
WebAssembly(\wasm)~\citep{haas2017bringing} is a binary format for a stack-based virtual machine with load/store architecture. It is designed to be compact, portable, secure, and efficient for web applications and beyond. It provides a coarse-grained memory isolation model where all application data are allocated in a contiguous \textit{linear memory} and all memory accesses are checked against the bounds of this linear memory, hence ensuring memory \textit{isolation}.  In addition, it utilizes type checking to ensure safety of the 
\wasm stack. It also provides lexical structure at the assembly level to 
prevent control-flow related attacks.

While \wasm prevents a buggy application from tampering memory of other applications using the linear memory sandbox, it does not stop one data structure to overflow another in the same insecure application. Recently, \textit{memory safe} \wasm~(\mswasm) was introduced, which is an extension of \wasm that provides this missing fine-grained memory safety protection~\cite{michael2023mswasm}.

\newcommand*{\ns}[1]{{n_{\texttt{#1}}}}
\newcommand*{\segid}{{n_{\texttt{id}}}}
\newcommand{\yl}[1]{\colorbox{yellow}{#1}}
\begin{figure}
    \caption{\centering \mswasm Syntax and Semantic of \wasm with \mswasm Extension highlighted~\citet{michael2023mswasm}. We add \textsf{h.null}, shown in pink, to the formal syntax to facilitate our proofs.}    \label{fig:mswasmsyntax}
        \begin{tabular}{rlllrlll}
            \textsf{Types}&        $\tau$~$\coloneqq$ &i32| i64 | f32 | f64  \yl{|~handle} 
            &\textsf{Value}& $v$~$ \coloneqq$ & ...~\yl{|~$\langle \ns{base}, \ns{offset}, \ns{bound}, b_{\texttt{valid}}, \segid \rangle$} \\
        \end{tabular}\\
        \begin{tabular}{rlll}
            \textsf{Instr.}& $i$~$\coloneqq$ & const $v$ |  ... | \yl{segload $\tau$|segstore $\tau$|~slice~|~segalloc~|~h.add~|~segfree}\colorbox{orange!30}{|~h.null} 
        \end{tabular}\\
        \begin{tabular}{rlrlrl}
            \textsf{Byte}&$b$~$\in \{0 .. 2^8 - 1\}$  &  
            \textsf{Lin. Mem}&$H$~$\coloneqq$ $b^*$ & \textsf{\yl{SegID}}&$\segid$~$\coloneqq$ i32  \\ 
            \textsf{Store}&$\Sigma$~$\coloneqq$  ($H$, \yl{$T$, $A$})   &
            \textsf{\yl{Segments}}&$T$~$\coloneqq$  i32 $ \hookrightarrow (b,t)$&  \textsf{\yl{Tags}}&$t$~$\coloneqq$ \textbf{H} | \textbf{D} \\ 
            \textsf{\yl{Allocator}}&$A$~$\coloneqq \segid \hookrightarrow$ (i32, i32) 
        \end{tabular}
    \end{figure}

\noindent \textbf{Syntax and Semantic.} \mswasm extends the \wasm syntax with six operations and a new type \textit{handle}, as shown in \Cref{fig:mswasmsyntax}. A handle is a new type that is essentially a fat pointer. A handle consists of four integers (a lower bound, an upper bound, a current offset, and an allocation ID), as well as, one validity byte, which denotes whether the handle is valid and active or not. Besides the normal pointer operations (e.g., pointer arithmetic) over handle,  \mswasm also defines a \textit{slice} operation to shrink the bounds of a handle. The handle structure together with its operations ensure complete spatial~(lack of overflow, underflow, etc.) and temporal~(lack of use after free, etc.) memory safety. When a violation happens, \mswasm semantics captures it and terminates. For ease of our formalization, we add ``h.null'' syntax, to represent a null handle.

\mswasm extends the \wasm semantic with the notion of \textit{segment} $T$ and \textit{segment allocator} $A$. A segment is a piece of memory disjoint from the linear memory in \wasm. Handles can only point to segments, and can also be serialized/deserialized in the segments. Semantically, every byte of every segment has a \textit{memory tag} that determines if it is data or handle, disallowing forging a new handle from data bytes.

\subsection{\checkedc}

\checkedc~\cite{ruef2017checked,Tarditi2023spec} is a dialect of C that adds spatial memory safety protection to C. A compiled \checkedc program is binary-compatible with other binaries on the same platform. \checkedc enhances C with a type system to declare a bound for a pointer, hence creating a \emph{checked} pointer. All checked pointers will be bound-checked for spatial safety at dereference time. For each bound check, the compiler will first try to prove that the check never fails and can be removed in binary; if not, it adds runtime checks that could be hoisted or optimized away by compiler optimizations. Therefore, we choose \checkedc as the backend of \aegis for its C compatibility and low-overhead spatial safety enforcement.

Pointer \lstinline{T* p} can be defined as a checked pointer pointing to an array as:
\begin{lstlisting}
array_ptr<T> p: count(n)
\end{lstlisting}

\noindent Where \lstinline{p} is a checked pointer of type \lstinline{T} whose lower bound is \lstinline{p} and whose upper bound is \lstinline{p + sizeof(T)  * n}. \lstinline{n} can be variable, \lstinline{struct} members, or an expression. Besides \lstinline{count}, \lstinline{byte_count} syntax can be used to specify the bounds size in bytes. Besides pointers to arrays, Checked C uses \lstinline{ptr<T>} for pointers to a single object~(i.e., address-of type pointers) that cannot have arithmetic, as well as, \lstinline{nt_array_ptr<T>} for pointers to null-terminated arrays.

\checkedc allows to gradually port C code to spatially-safe, typed \checkedc code. A piece of code can be marked as ``checked'' or ``unchecked'' at different levels of granularity. \checkedc's Blame Theorem guarantees that if a piece of code is checked, then that piece of code cannot be ``blamed'' for any spatial safety violation~\cite{ruef2017checked}. When there is a mix of checked and unchecked code, it is not possible, in general, to guarantee spatial safety.
Our tool, by virtue of \mswasm's semantic being spatially-safe, automatically generates checked code from the original C code, except for variadic functions that are currently not supported in \checkedc and outgoing calls to those external C functions that we cannot analyze or instrument. Thus, our tool generates spatially-safe \checkedc code from \mswasm code, subject to its limitations.

\section{Formal Semantics of \mswasm in Coq}\label{sec:formal}
Semantics of \mswasm is proven memory safe~\citet{michael2023mswasm}, but not mechanically using a formal tool. We have developed a mechanical formal semantics of \mswasm together with formal proofs about its isolation guarantees. Our formalization is based on WasmCert-Coq \cite{watt2021two} and Iris-Wasm \cite{rao2023iris}. In the rest of this section, we briefly explain our results, focusing on module systems of \mswasm and how its semantics correspond to our main result on \mswasm isolation guarantee. We decorate Coq formalized definitions, theorems, or examples with \coqverified. We provide our full formalization.

\subsection{Extension for \mswasm on Module Interaction}
We first briefly recall the \wasm module design in \Cref{fig:mswasmmodulesyntax}. A module has four components: functions $f^*$, linear memory $mem^*$, (function) tables $tab^*$, and global variables $glob^*$. A \wasm module will be \textit{instantiated} to an \textit{instance} as its runtime representation with these four components.  The global variables are statically typed, like \wasm stacks. Each component can be imported, exported, or initialized during (MS)WASM \textit{module instantiation}. Once the module instantiates, the starter function will be automatically invoked if there is one.

\begin{figure}
  \caption{{(MS)WASM Module Syntax, adapted from \citet{rao2023iris}}\label{fig:mswasmmodulesyntax}}
  \begin{tabular}{rlllrlll}
      \textsf{(glob. ty.)}&        $tg$&$\coloneqq$ &
      \textbf{mut}~$\tau$ | \textbf{immut}~$\tau$
      
        &\textsf{(import type)}&        $imptdes$&$\coloneqq$ &
        $\textbf{func}_i$ ~n | ... | $\textbf{glob}_i$ $tg$
        \\
        \textsf{(import)}&        $impt$&$\coloneqq$ &
        \textbf{import}~$imptdes$ 
        &
        \textsf{(export type)}&        $exptdes$&$\coloneqq$ &
        $\textbf{func}_e$~n | ... | $\textbf{glob}_e$~n
        \\
        \textsf{(export)}&        $expt$&$\coloneqq$ &
        \textbf{export}~$exptdes$ 
        &
        \textsf{(globals)}&        $glob$&$\coloneqq$ &
        \textbf{global}~$tg$~$i$ 
        \\
    \end{tabular}
    \begin{tabular}{rlll}
          \textsf{(modules)}&        $m$&$\coloneqq$ &
          \textbf{module}~$f^*$~$glob^*$~$tab^*$~$mem^*$  $impt^*$~$expt^*$
    \end{tabular}
\end{figure}

The \mswasm original paper~\cite{michael2023mswasm} considers the scenario where only one module is active at a time, and thus the imports and exports are excluded from their formalization and implementation. We present our formalism that includes multiple modules. Compared to \wasm, \mswasm allows a module to import and export a handle, but it is important to notice that a handle cannot be stored in the linear memory, and thus the only way to import and export a handle is to make it a global variable. %To initialize global handle variables during module instantiation, we added the new instruction \texttt{h.null} which is not formally mentioned in \citet{michael2023mswasm} but already implemented in the \mswasm artifact.

Our isolation proof for \mswasm is defined incrementally over the structure of \mswasm syntax, using the method of \textit{logical relation}~\cite{dreyer2018type}.

\subsection{Type Safety in \mswasm}
This is the first step in our incremental proof. We extend the Iris-Wasm's unary logical relation with \mswasm features and derive a strengthened type safety compared to \cite{michael2023mswasm,watt2021two}: arbitrary syntactically well-typed \textit{partial} programs can execute successfully after proper linking. Then we can derive the type safety for \textit{complete} programs.

\begin{cthm}[Type Safety for Closed Expression]  
    A syntactically well-typed closed expression can never reach a \textit{stuck state} during its execution, where stuckness represents any error including spatial safety. 
\end{cthm}

\noindent Thanks to the design of \wasm, this type safety already covers the safety of the \mswasm stack. Please note that \texttt{trap} is a well-defined behavior in \wasm, that is not considered ``stuck''. 

Next, we lift type safety to the module level, based on the module level formalism in Iris-Wasm.

\begin{cthm}[Type Safety for Closed Module]
    A syntactically well-typed closed module (with a starter function) can never reach a stuck state during its execution.
\end{cthm}

\newcommand*{\reachable}[2]{{\texttt{reachable}(\ensuremath{#1}, \ensuremath{#2})}}
\newcommand*{\Reachable}[2]{{\texttt{Reachable}(\ensuremath{#1}, \ensuremath{#2})}}

\subsection{Memory Safety in \mswasm}
\citet{michael2023mswasm} provides a memory safety proof based on a \textit{monitor machine}, covering the spatial and temporal safety of the \mswasm. However, this safety property cannot capture the \textit{isolation} and \textit{handle integrity} properties relevant for \mswasm: it is possible for a spatial and temporal safe language to allow forging handles. We illustrate how our mechanization can help fill this gap and strengthen the formal guarantee of \mswasm.
First, we need some terminology.

\begin{defn}[\textit{Reachable}, and \textit{Maximal Reachable}]
  We say a segment $S$ is reachable from an active handle $h$, denoted as a relation \reachable{h}{S} when  
  \begin{itemize}
    \item we can access $S$ from handle $h$
    \item if handle $h$ can access another handle $h'$ and \reachable{h'}{S'}, then \reachable{h}{S'}
    \item \reachable{h}{S} and \reachable{h}{S'} implies \reachable{h}{S \cup S'}
  \end{itemize}

\noindent  We define the maximal reachable \Reachable{h}{S} as the largest $S$ s.t. \reachable{h}{S} holds.
\end{defn}

\begin{cthm}[Local Handle Isolation]\leavevmode\label{Thm:LHE}
\begin{itemize}
  \item Given arbitrary \textit{syntactically well typed} function $f$,
  \item given a set of well-typed arguments $vs$ for $f$,
  \item if we can separate the whole memory $M = L \dot{\cup} E$ into two disjoint parts, \\ where we call $L$ \textit{local memory} (or \textit{non-exported memory})  and $E$ \textit{exported memory},
  \item if \Reachable{vs}{E}
\end{itemize}

\noindent  then during and after the function call $f(vs)$, the local memory $L$ is unchanged.
\end{cthm}

\noindent In above theorem, \textit{exported memory} is defined as the (maximum) reachable memory from the function inputs $vs$; and \textit{local memory} is just the rest of the memory. 
Our theorem can explicitly show the invariance of the \textit{local memory} during the execution of $f$. In other words, this property means ``\textit{the execution of $f$ \textbf{will never write} the non-exported memory}''.

Another way to look at Theorem~\ref{Thm:LHE} is to consider it  as a \textit{safety property}~\cite{lamport2002specifying}, because its violation can be detected by monitoring modifications in {local memory} in a finite trace. More importantly, this safety is \textbf{robust} against arbitrary (even malicious) linking. We demonstrate this using a concrete example (the theorem holds for all cases).

% add example
% []
\begin{cexample} Given an arbitrary \textit{syntactically well-typed} function $\vdash g : [] \to []$ (an arbitrary function representing an ``attacker''), and a parameter $n$, we have the following code:
\begin{lstlisting}[language=MSWasm]
(const $4$); (segalloc); (local.tee $0$);    ;;Allocation;;
(const $n$); (segstore i32);                 ;;Store n;;
(call $g$);                                  ;;Hand over to attacker ;; 
(local.get $0$); (segload i32);              ;;Retrieve the stored value;;
(const $n$); (sub i32); (eqz i32);           ;;Check Modification;;
(if [const 1] [const 0])                  ;;Return 0 if modified, otherwise 1;;
\end{lstlisting}

Basically, we allocate local memory to store $n$, and we check if this local memory has been modified after the execution of the attacker, $g$.

Using our mechanical formalization and proof, we can formally show that this piece of code will either terminate with 1 or trap (during the execution of  $g$). That means an arbitrary attacker, $g$, cannot modify the local memory of the caller. 
\end{cexample}

This mechanized example further illustrates the \textit{robustness} and \textit{isolation} against arbitrary (well-typed) attackers. 
A corollary of Theorem~\ref{Thm:LHE} is \textit{handle integrity}: it is not possible for an attacker $g$ to forge the local handle in the caller, or else it could have consequently modified the caller's memory. 
We also accompany our mechanization with several other examples, including those involving modules, to illustrate the usage of our logical relation model.

Finally, we can lift Theorem~\ref{Thm:LHE} to the module level, by considering the imports and exports as inputs and outputs.
In this theorem, we will ignore the import of the table and linear memory, and focus on the impact of the segments. Thus, we need to extend the definition of reachable for function closures. The details can be found in the mechanized proof.  
\begin{cthm}[Local Handle Isolation for Module]\leavevmode
  \begin{itemize}
    \item Given an arbitrary \textit{syntactically well typed} module $F$ requiring no imports on linear memory and tables,
    \item given a set of well-typed arguments $vs$ as imported global variables
    \item and a set of closures $cs$  as imported functions,
    \item if we can separate the whole memory $M = L \dot{\cup} E$ into two disjoint parts, \\ where we call $L$ \textit{local memory} (or \textit{non-exported memory})  and $E$ \textit{exported memory},
    \item if \Reachable{vs}{E_1} and \Reachable{cs}{E_2} and $E = E_1 \cup E_2$
  \end{itemize}
  
\noindent    then during and after the module instantiation $F$ (including the execution of its starter function) using imported $vs$ and $cs$, the local memory $L$ is unchanged.
\end{cthm}

\section{\aegis Pipeline}\label{sec:impl}
Section~\ref{sec:formal}  highlights the protection guarantees of \mswasm. In this section, we present the design and implementation of our \aegis pipeline that preserves the formalism's protection guarantees when it comes to spatial safety. \aegis pipeline implements the spatial memory protection for the input program by generating code in \checkedc and resuming the build process using the modified \checkedc compiler, which can compile both C and \checkedc code. Compared to other memory safety schemes using runtime checking, \aegis has a special focus on preserving memory safety formal guarantees. Compared to existing C to \checkedc solutions that convert raw pointers to checked pointers in \checkedc, \aegis pipeline converts them into \emph{fat pointers} instead, losing the source code compatibility and efficiency in exchange for complete automation and coverage. We make our entire implementation available.

\begin{figure}[ht]
\centering
\includegraphics[width=\linewidth]{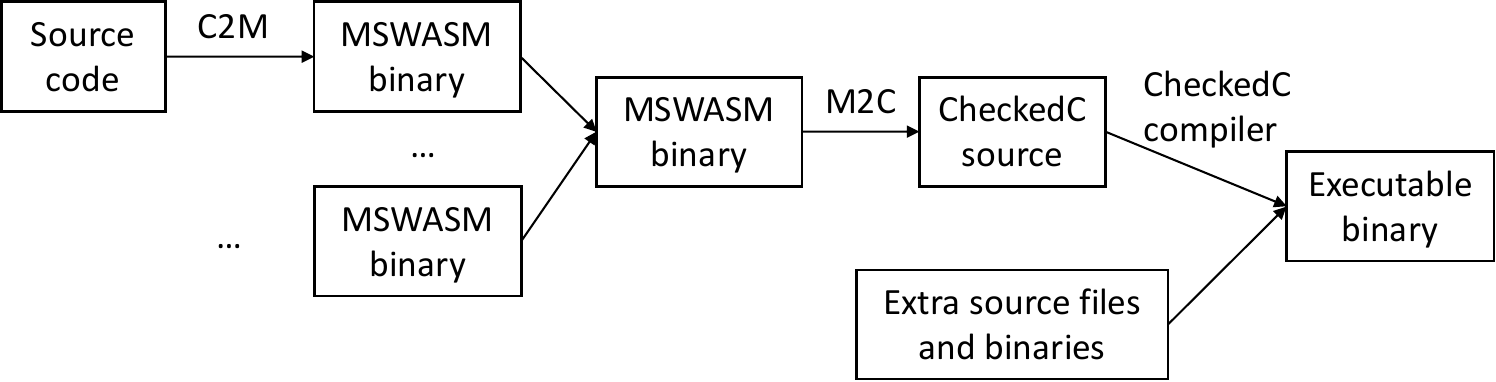}
\caption{Overview of AEGIS Pipeline}\label{fig:aegis_pipeline}
\end{figure}

Figure~\ref{fig:aegis_pipeline} gives an overview of the pipeline. First, the user would use a C-to-MSWASM~(\ctomswasm) compiler to produce the MSWASM binary. The original \ctomswasm compiler~\citet{michael2023mswasm} does not emit correct code for C programs using global variables and other unsupported features; they should be compiled with our modified \ctomswasm. The output MSWASM binaries are the counterparts of object files, and if there are multiple source files, all MSWASM binaries are linked together\footnote{Currently, \mswasmtocc requires the input MSWASM files to remain relocatable (i.e., preserving the ``linking'' section) if they depend on symbols from native code (e.g., libc).}. Next, the user runs our MSWASM-to-Checked-C code generator (\mswasmtocc) that emits \checkedc sources and headers from \mswasm files. After that, the user can use the \checkedc compiler to compile them as regular (Checked) C programs and resume normal build procedure. Each of these tools can be used separately; for example, \mswasmtocc can be applied independently to a valid \mswasm module. If there are additional source code and binaries, the user can add them when invoking the \checkedc compiler.

This pipeline allows C programs to benefit from the \mswasm as a secure bytecode format without losing C interoperability. The user can guarantee that all code in the input \mswasm files are spatially protected, even if some \mswasm files are produced by third party vendors. This property is especially useful when there are multiple code owners/vendors. Assume Alice develops one component of a system and delivers it to Bob, who in turn collects binaries from multiple parties, integrates them together, then ships the executable to customers. 
Previously, Alice ships native binaries to Bob, and Bob cannot verify the security of the binary unless there is some non-trivial binary analysis in place. With the \aegis pipeline, Alice can ship \mswasm binaries to Bob, and Bob can compile this binary into native code with confidence that the delivered code has spatial memory safety protection.

Besides converting the entire program, \aegis also supports protecting part of the program (e.g., a vulnerable library) by selectively compiling the selected files through \aegis pipeline and writing wrappers for functions defined in them. Writing wrappers that export \aegis-protected functions to native C code still requires some programmer effort at the time of writing.

In summary, our implementation includes:
\begin{itemize}
\item Modifying  \ctomswasm to support various missing features, including C global variables and fine-grained spatial safety for stack via ``slice'' instruction in \mswasm; and to fix bugs;
\item Creating \mswasmtocc code generator, which is based on the \texttt{wasm2c} tool in the WABT from \mswasm fork but with the code generation section rewritten; and lastly
\item Modifying \checkedc compiler, mainly, to support customizable error handling upon spatial safety violation.
\end{itemize}

\paragraph{Limitations}
Currently, function pointers in native C and \mswasm has different representations and thus cannot be used across the domains. 
\aegis does not support programs passing pointers/handles between native C and \mswasm through shared memory in the general case,
and requires programmer effort to write wrappers. 
However, \aegis automatically converts pointers passed through function call/returns in best efforts.

\subsection{\aegis Pipeline In Action}
Next, we use a descriptive example of how \aegis works. We have also developed a pen-and-paper formalization for proving the correctness of our \mswasmtocc; a technical report for this proof is included in our provided artifacts.

\begin{figure}[ht]
    \centering
    \includegraphics[width=\linewidth]{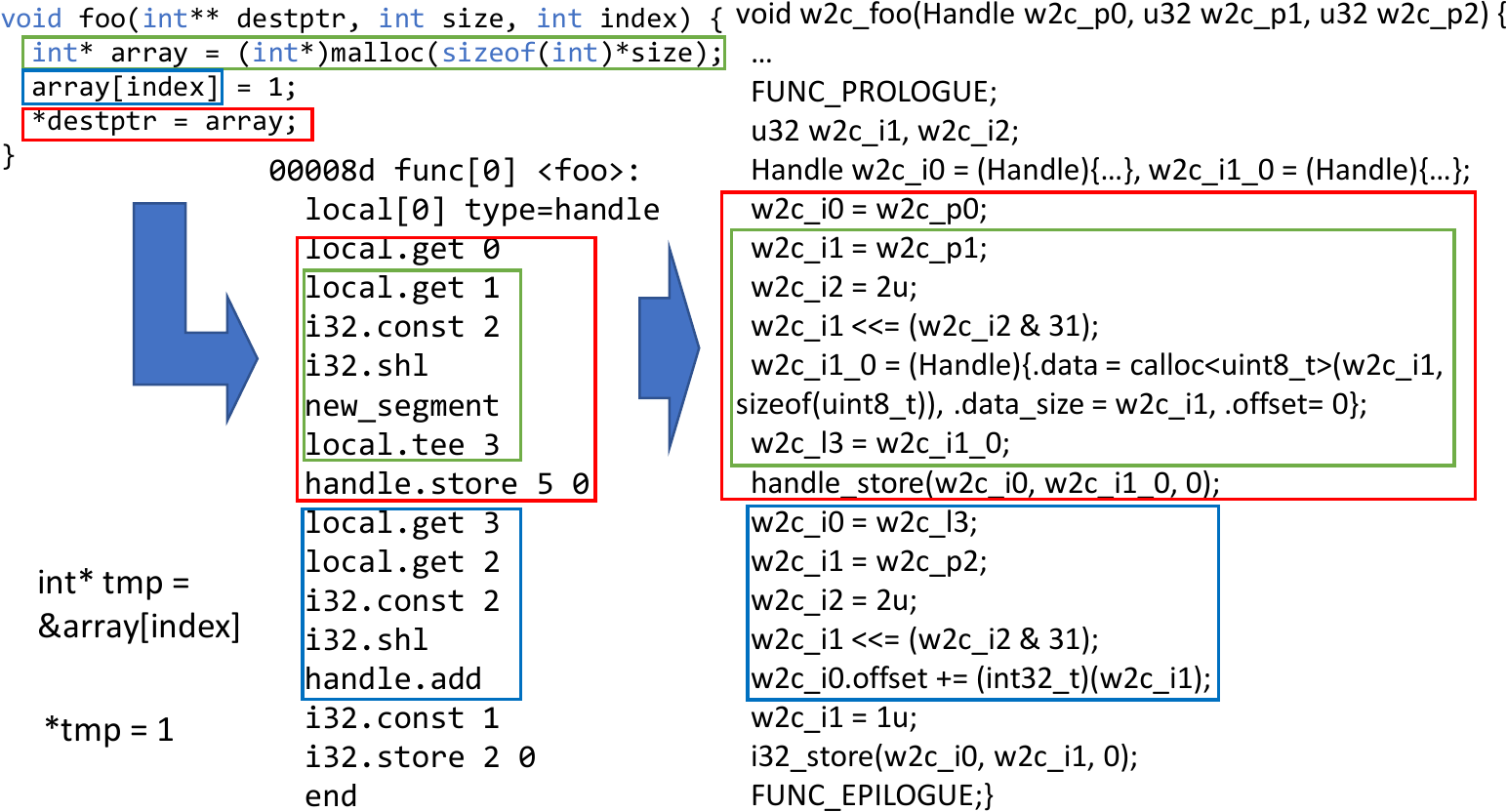}
    
    \caption{Code example (C to MSWASM to Checked C)}\label{fig:impl_aegis_code_1}
\end{figure}

Figure~\ref{fig:impl_aegis_code_1} shows a simple example of how \aegis pipeline converts C code to \mswasm and then to \checkedc. The green, blue, and red boxes demonstrate how different parts of original C code are transformed into \mswasm, and \checkedc. As a stack machine, \mswasm instructions consume values on the stack and push results back to the stack. We annotate the sub-expression the code is evaluating for the blue box (\lstinline{&array[index]}) and code after it (storing value 1 to the address), for readability. \mswasm maps function arguments to local variables, and \lstinline{local.get 0}/\lstinline{1} in \mswasm pushes \lstinline{destptr} and \lstinline{size} in C, correspondingly. In the green box, MSWASM converts the \lstinline{malloc()} to \lstinline{new_segment} that pops a size and pushes the resulting handle into the stack. The three instructions before \lstinline{new_segment} is computing \lstinline{size} left shifted by 2 (which is \lstinline{sizeof(int)*size}). The \lstinline{handle_store} after them writes the \lstinline{array} pointer to \lstinline{destptr}. The code inside blue boxes computes the address of \lstinline{array[index]}. \lstinline{local.get 3} pushes \lstinline{array} to the stack. Instructions from \lstinline{local.get 2} to \lstinline{i32.shl} compute \lstinline{index*sizeof(int)} as the offset for the pointer arithmetic. \lstinline{handle.add} adds them together. Finally, the code after the blue boxes dereference the resulting pointer and stores a constant value \lstinline{1} to it.

\mswasmtocc translates \mswasm functions similar to macro expansion: each instruction maps to a \checkedc code sequence implementing its semantics, and \mswasmtocc uses a big switch statement inside a loop that pattern-matches instructions to convert each function. We also rewrote code sequences to replace the use of \wasm linear memory with \mswasm's segmented memory, as linear memory is represented as segment itself in \mswasm semantics.

\begin{lstfloat}[ht]
    \centering
\begin{lstlisting}
typedef struct Handle{ // defined in runtime library
  long data_size; // n_bound, max size
  long offset; // n_offset, current offset from base address
  _Array_ptr<uint8_t> data: count(data_size); // n_base, base address
} Handle; // T* ptr, *ptr --> Handle h, *((ptr<T>)(h.data + h.offset)

static inline void handle_store(Handle h, Handle value, int offset) {
  dynamic_check(sizeof(Handle) <= h.data_size);
  dynamic_check((unsigned long)(h.offset + offset) <=
    (unsigned long)(h.data_size - sizeof(Handle)));
  *((ptr<Handle>)(h.data + h.offset + offset)) = value;
}
\end{lstlisting}
    \caption{Handle definition and handle\_store() Implementation}\label{lst:impl_aegis_code_2}
\end{lstfloat}

Listing~\ref{lst:impl_aegis_code_2} shows how \mswasmtocc implements \mswasm handle and memory accesses. We implement \lstinline{Handle} as a struct with a checked pointer and bounds information.

We implement all memory access related instructions using handles, for example, dereferencing the handle \lstinline{h} to type \lstinline{T} becomes \lstinline{*((ptr<T>)(h.data + h.offset))}. There are various load and store operations in \mswasm, which are implemented similarly. For example, \lstinline{handle_store(h, v, offset)} in Listing~\ref{lst:impl_aegis_code_2} stores the value of handle \lstinline{v} into where handle \lstinline{h} points. Because (MS)WASM memory access instructions take an additional \lstinline{offset} value, the code adds this \lstinline{offset} as well when dereferencing a handle. Because all generated \checkedc code is in the checked scope, the \checkedc compiler will either try to prove that the access is statically safe, or it will insert runtime bounds checking to guarantee spatial safety. The \checkedc \lstinline{dynamic_check()} calls express necessary and sufficient conditions for spatial safety. If \checkedc compiler can prove that a \lstinline{dynamic_check()} condition always holds, then it is removed in runtime, otherwise the compiler will insert runtime checks to enforce it.

When the \mswasm program contains calls to functions defined outside, \mswasmtocc assumes the callees are native C functions and generates \emph{wrapper} functions to handle calling convention differences between \mswasm and native C. The wrapper functions convert C pointers and \mswasm handles for pointer arguments and return values, and recovers struct's passed by value
in generated \checkedc source. Because C pointers do not contain bounds information, the current implementation creates handles with infinite bounds, bypassing bounds checks on them. Our implementation allows users to manually prepare the wrappers so that tighter bounds can be specified using human knowledge. The wrapper functions are defined with weak linkage so that if another \mswasm module has a strong definition, it takes priority during linking.

\subsection{Additional Features}
Besides the \mswasmtocc code generator, we modified \ctomswasm and \checkedc compiler to support several language features and use cases while observing the MSWASM semantics in the pipeline.

\subsubsection{Global variable handling}
While (MS)WASM has ``globals'', they cannot be used for C global variables because they are addressed by indices and it is not possible to take their address. Therefore, in the (MS)WASM implementation, global variables and constant data (e.g., constant string literals) are compiled to ordinary data symbols in a single data segment reserved for them. However, their initializers are emitted as WASM data segments, which, in the MSWASM formalism, should be mapped to WASM linear memory. In addition, accesses to global variables are not spatial-safety protected because they do not use handles.

In our implementation, we changed the code generation for global variables in the \ctomswasm compiler so that: (1) a backend-specific IR pass creates a new \mswasm global handle for each global variable declaration and emits initialization functions to initialize the handle; and (2) the Instruction selection (ISel) lowers all global variable references as loading and dereferencing the global handle. The initialization functions will be executed once during the module initialization according to WASM semantics. In addition, if \mswasmtocc finds such a global handle declaration in the input MSWASM binary, it will emit C global variable declarations or definitions in the output \checkedc code, and initializes the handles with the address of the global variables. This ensures that the C source code in \aegis pipeline can reference global variables defined outside and vice versa.

\subsubsection{Bounds narrowing}
\mswasm includes the slice instruction, with the usual semantics of returning a sub-handle from a given handle. However, the instruction is not implemented in the original \mswasm toolchain. In addition, utilizing the slice instruction for spatial protection requires computing the change to lower and upper bounds, which requires a sequence of instructions to implement. This would result in code bloat and it is inefficient in practice. Instead, we adopt CHERI's \texttt{CSetBounds} instruction semantics and implemented \texttt{handle.setbounds} instruction that takes (1) the handle for bounds narrowing, and (2) the new length that the handle is allowed to access. This instruction can be shown to be equivalent to the semantics of slice in \mswasm. This instruction allows us to protect access to stack objects: when the address of an stack-allocated object is taken, we emit a \texttt{handle.setbounds} instruction so that the result pointer carries bounds to cover only the referenced object instead of the entire stack. This new feature provides spatial safety for stack.

\subsubsection{Integer-to-pointer cast}
C programs written for embedded systems sometimes need to read or write at specific hard-coded addresses for I/O or other low-level system operations. These accesses becomes integer-to-pointer cast followed by pointer de-references in the LLVM IR. To support these programs while maintaining the formalism of MSWASM, the MSWASM module imports a special handle \texttt{\_physical\_memory}\footnote{The mechanism is not limited to bare-metal programs; it can be used in virtual address space} that points to a segment representing the underlying address space, and integer-to-pointer casts are lowered as pointer arithmetic on this handle. We modified \mswasmtocc and \checkedc compiler as well to support recovery of such operations on access to \texttt{\_physical\_memory}.

\subsection{Case Studies}

We have applied our toolchain to secure real-life vulnerabilities. We have used our tool in both: (i) ``whole-program'' mode, where all source codes are transformed into \checkedc code; and (ii) ``partial-program'' mode where only a selected part of a system (e.g., a file) is transformed. We have also created a cross-compilation toolchain that enabled us to port our toolchain to an embedded domain.

We now briefly discuss how we used \aegis to treat two well-known critical spatial safety vulnerabilities: 
(i) Heartbleed vulnerability in OpenSSL~(CVE-2014-0160); and (ii) \texttt{git log} vulnerability in Git~(CVE-2022-41903). 
Both vulnerabilities happen on the heap, due to missing bound checking in \lstinline{memcpy} calls. The former is a buffer overflow vulnerability. 
The latter is a buffer underflow vulnerability, itself the result of integer overflow.\footnote{https://www.x41-dsec.de/static/reports/X41-OSTIF-Gitlab-Git-Security-Audit-20230117-public.pdf} In both cases, we use \aegis selectively over the vulnerable parts of the code. 
The rest of the code communicates with the transformed code via wrappers that the tool provides which we customize. The main effort went into parameters of the vulnerable function that included pointers to structs that needs to be serialized/deserialized to the 
corresponding handle generated in \aegis \checkedc. While \aegis automatically generates wrappers for interfacing, 
it is necessary to manually specify bounds for raw C pointers which are converted to handles in \aegis code.
We also make minor modifications to include \aegis transformed code in the build systems of OpenSSL and Git. 
In both cases, our fix effectively secured the code against the known vulnerabilities, with little effort. 

The mechanism by which \aegis secures the faulty \lstinline{memcpy} calls is straightforward: A pointer in the original C code is eventually transformed into an \mswasm handle, and subsequently a \checkedc pointer, which guarantees to capture any access beyond its static low and high bounds. In \aegis, we have two options for implementation of secure \lstinline{memcpy}:

\begin{itemize}
\item The libc implementation of \lstinline{memcpy} can be compiled via \ctomswasm and \mswasmtocc to produce an equivalent version of \lstinline{memcpy} that accepts handles:
\begin{lstlisting}[title={\lstinline{memcpy} prototype via Handles}, 
      frame=tlrb, label={list:memcpy_inline}]{memcpy_inline}
      Handle w2c_memcpy(Handle w2c_p0, Handle w2c_p1, u32 w2c_p2)
\end{lstlisting}

\item The call to \lstinline{memcpy} is considered an external call to a libc function by \aegis, allowing us to provide a \checkedc interface of \lstinline{memcpy} that accepts checked pointers instead of raw pointers:

 \begin{lstlisting}[title={\lstinline{memcpy} prototype via Checked C pointers}, frame=tlrb]
itype_for_any(T) void *memcpy(
	void * restrict dest : itype(restrict array_ptr<T>) byte_count(n),
	const void * restrict src : itype(restrict array_ptr<const T>) byte_count(n),size_t n) : itype(array_ptr<T>) byte_count(n);
\end{lstlisting}
\noindent where the above prototype in \checkedc asserts that the source and destination buffers can be checked pointer whose sizes are bound by the last parameter of function, \lstinline{n}.  The call goes through WASI interface.\footnote{https://github.com/WebAssembly/WASI} 

\end{itemize}

\noindent Both approaches work in \aegis, with the second option being faster but would require the \mswasmtocc compiler to provide the proper checked pointers for \checkedc version of \lstinline{memcpy}.

\section{Evaluation}\label{sec:eval}
We evaluate our \aegis pipeline to measure its impact on the performance and binary size of the protected programs. We use PolyBenchC \footnote{https://github.com/MatthiasJReisinger/PolyBenchC-4.2.1} for evaluation, which is commonly used for evaluating \wasm compilers. This benchmark is also used by the original \mswasm paper when evaluating their Rust backend for \mswasm~\cite{michael2023mswasm}, which itself is based on rWasm \cite{bosamiya2022provably}. We run all programs on a Ubuntu 22.04.2 system with an Intel i9-9900K CPU. We compare the protected versions of programs generated by the following three tools against native:

\begin{enumerate}
  \item \aegis pipeline: We use \ctomswasm (Clang version 11.0.0 with -O3), \mswasmtocc and \checkedc (Clang version 12.0.0 with -O3).
  \item $rWasm_{bb}$: We use the Baggy Bounds \cite{akritidis2009baggy} implementation of rWasm, which converts \mswasm to Rust \cite{michael2023mswasm}.
  This technique stores memory as a single vector of bytes, and an allocator allocates segments which allow for faster bound checking. 
  Handles are 64-bit values storing an offset in memory and the log of the segment size (rounded up to nearest power of two at allocation). 
  \item $rWasm_{seg}$: We use the segments implementation of rWasm \cite{michael2023mswasm}. 
  This stores segment memory as a vector of segments, where each segment is a vector of bytes, thus automatically enforcing spatial safety, thanks to built-in spatial safety in safe Rust. However, this implementation also provides temporal safety - we manually turn off this feature to accurately measure overhead for only spatial safety.
\end{enumerate}

\subsection{Performance Impact}
\begin{figure}[ht]
\centering
\subfloat[Normalized runtime\label{fig:eval_perf_runtime}]{\includegraphics[width=\linewidth]{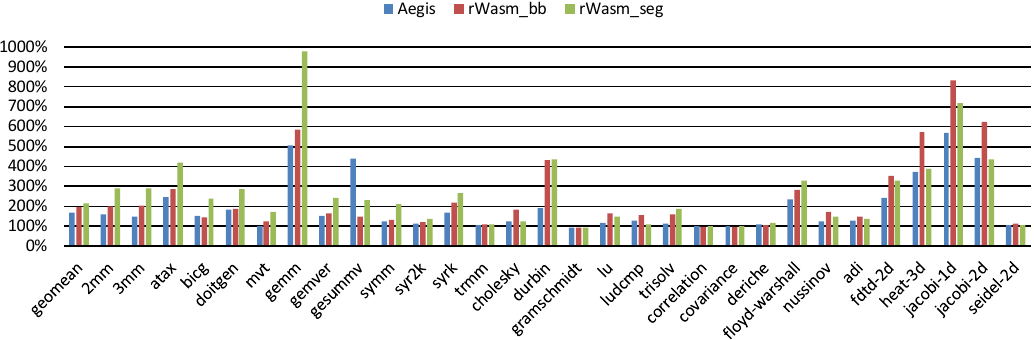}}

\subfloat[Peak memory usage (KB), stacking \textbf{V}irtual (dark) with \textbf{P}hysical (light)\label{fig:eval_perf_memory}]{\includegraphics[width=\linewidth]{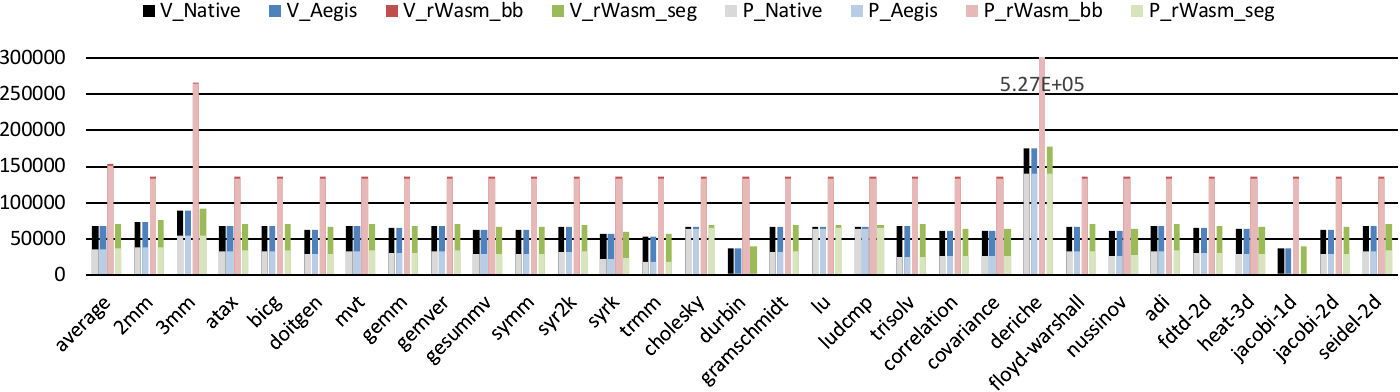}}
\caption{Runtime and memory usage.}\label{fig:eval_perf}
\end{figure}
\begin{figure}[ht]
\centering
\includegraphics[width=0.5\linewidth]{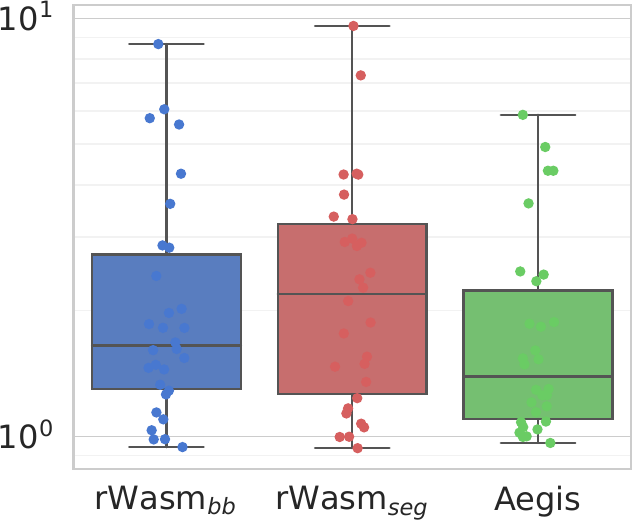}
\caption{Normalized slowdown against native (log scale); cf. Figure 13 in~\cite{michael2023mswasm}.}\label{fig:eval_runtime_logscale}
\end{figure}

We compare the runtime and memory usage of PolyBenchC programs protected by the three tools above. To evaluate the runtime increase, we use PolyBenchC's builtin timer, which outputs execution time using \lstinline{gettimeofday} function. Therefore, the runtime numbers exclude startup and shutdown phase of the programs. This decision may make the reported runtime overhead smaller than the end-to-end measurement, but it facilitates more fair comparison among the tools we evaluate. To measure the memory usage, we retrieve the peak virtual (\texttt{VmPeak}) and physical (\texttt{VmHWM}) memory usage from Linux \texttt{proc} file system. Figure~\ref{fig:eval_perf} shows the runtime and memory usage of each versions.

Figure~\ref{fig:eval_perf_runtime} shows the normalized runtime usage. The line for 100\% represents the native execution, and bars above 100\% represent the overhead portion of the results. When compared to native, \aegis has 67\% geomean overhead, $rWasm_{bb}$ has 97\% overhead, and $rWasm_{seg}$ has 117\% overhead. \aegis has the least overhead, thanks to \checkedc compiler optimizations and our efficient implementation of \mswasm runtime. $rWasm_{bb}$ has better performance compared to $rwasm_{seg}$ since $rWasm_{bb}$ uses the Baggy Bounds technique which has a reduced cost for bounds checking at the cost of higher memory consumption.

Figure~\ref{fig:eval_perf_memory} shows the virtual and physical memory usages. \aegis incurs near-zero memory overhead in all programs because it requires no changes on object allocations and no extra dependency. $rWasm_{seg}$ incurs 5\% geomean overhead compared to native execution because of the Rust implementation. $rWasm_{bb}$ has the highest memory overhead, ranging from 87\% to 272\% compared to native for virtual memory usage and 109\% to 6430\% for physical memory. This is because it allocates objects in power-of-two sized bins.

\subsection{Executable Size}
Because PolyBenchC programs have similar code sizes and all native executables are between 17KB and 21KB, we discuss average executable sizes only. Executables produced by the \aegis pipeline is 29KB on average, 70\% larger than the native versions. In contrast, 
the average executable sizes for $rWasm_{bb}$ and $rWasm_{seg}$ are 35,375KB and 35,650KB respectively. This massive code bloat is from the libraries Rust statically linked into the executables.

\section{Related Work}\label{sec:related}

\subsection{Pointer-based Spatial Safety Protection}
Fat pointer schemes, such as \aegis, store a pointer and its bounds next to each other to achieve locality, and thus efficiency. CHERI~\cite{cheri} architecture introduces dedicated capability registers and instructions to accelerate fat pointer operations. 
\mswasm is designed with CHERI backend in mind. Besides storing bounds inline with pointers for efficiency, schemes can also store them out-of-band for better compatibility with uninstrumented binaries~\cite{mpxexplained,softbound}. Fat pointer schemes can enforce fine-grained spatial safety at the cost of compatibility.

To overcome the incompatibility of memory layout by fat pointers, early works store the pointer metadata in disjoint memory locations (i.e., shadow memory)~\cite{10.1145/1542476.1542504,watchdoglite}, incurring high performance overhead. More recent work on tagged pointers squeeze metadata into the unused bits of pointers. This enables embedding metadata without changing pointer size or paying pointer metadata lookup cost.
FRAMER~\cite{framer} and In-Fat Pointer~\cite{ifp} use the pointer tags to locate in-memory object bounds that are larger in size.
Tagged pointer schemes achieves better compatibility at the cost of weaker security guarantees.
Further work related to memory safety using dynamic checking can be found in \cite{sanitizers,memsaneval}.

\subsection{Transforming C code to Safer Languages}
With the advent of memory-safe programming languages, researchers also started exploring source-to-source transforms that port unsafe C programs to safer languages. However, because of the semantic gap between C and safe languages, it is challenging for these analyses to be sound and complete on potentially unsafe operations. Therefore, these transforms are not fully automatic or they don't have full coverage yet.
3C~\cite{machiry2022c} introduces progressive protection to C code by inferring Checked C type annotations that place safety checks.
CheckedCBox~\cite{checkedcbox} applies program partitioning to Checked C programs to protect checked code from unchecked code. There are also works exploring converting C to Rust programs using static analysis~\cite{c2rust_aliasing,c2rust_ownership}.

\subsection{Formal Model on Heap Safety}

\noindent \textbf{1-Safety on heap.} SoftBounds~\cite{10.1145/1542476.1542504} and \checkedc~\cite{li2022formal} provide extensions on the formal semantic of a fragment of C in Coq and prove their (syntactic) type safety. SoftBounds focuses on the absence of over-bound, while \citet{li2022formal} aims at \textit{The Blame Theorem} to characterize Core \checkedc as a gradual type system.

RichWasm~\cite{paraskevopoulou2024richwasm} extends single-module \wasm semantic with a substructural capability-based type system and proves its (syntactic) type safety. With the advanced type system design, spatial and temporal safety can be eliminated in RichWasm at compile time, at the cost of expressing potentially unsafe code.

\noindent \textbf{Non-interference on heap.} The closest work is \citet{de2018meaning}, which argues the definition of memory safety should include aspects of privacy and isolation, and consequently a form of non-interference.

Another stream of works on non-interference is based on \textit{Information Flow Type System} \cite{frumin2021compositional,barthe2007certified,kammuller2008formalizing}. These works aim at a more general setting of non-interference---not restricted to heap---but require modifications to syntax and typing rules to additionally annotate security levels. %In contrast, \mswasm treats handles as capabilities just like \citet{de2018meaning} and thus can provide an intuitive concept of isolation and non-interference without the help of security level.

Compared to non-interference, our notion of \textbf{robust} 1-safety is weaker, as our final theorem can only claim no modification (no-write) happens in the designated area, but we cannot prove the absence of information leakage (no-read). However, our advantage is the simplicity of the proof structure, as it is a relatively lightweight extension to the existent unary logical relation in \citet{rao2023iris} and thus leaves most existent proof unchanged.

\section{Conclusion}\label{sec:conclusion}
We present the first mechanized formalization of \mswasm and extend it to support secure inter-module interaction. Based on our formalism, we also present \aegis, which provides a practical solution to apply \mswasm on C programs, by translating \mswasm into spatially-safe \checkedc. By virtue of generating code in a dialect of C, \aegis allows for a programmatic, binary-compatible level of interoperability between memory-safe Checked C and C code. Our formalization and implementation provide the foundation for adopting \mswasm in realistic settings where memory safety, interoperability, and performance are paramount.

\bibliographystyle{splncs04}
\bibliography{references}
\end{document}